\documentclass[12pt,preprint]{aastex}

\usepackage{amsmath,amssymb}
\usepackage{graphicx}
\usepackage{natbib}
\def\grm{{\tt grmonty}}
\def\mt{\,GM/c^3}
\def\ml{\,GM/c^2}
\def\msun{\,\rm M_{\odot}}

\def\mum{\,\mu\textrm{m}}

\def\imin{\,\rm min^{-1}}

\def\kev{\,\textrm{keV}}

\slugcomment{Accepted by ApJ Letters}
\shorttitle{QPOs from simulations of Sgr A*}
\shortauthors{Dolence et al.}

\begin{document}
\title{Near-Infrared and X-ray Quasi-Periodic Oscillations 
in Numerical Models of Sgr A*}
\author{Joshua C. Dolence\altaffilmark{1}, Charles F. Gammie\altaffilmark{2}, Hotaka Shiokawa}
\affil{Astronomy Department, University of Illinois at Urbana-Champaign, Urbana, IL, 61801}
\altaffiltext{1}{Current address: Department of Astrophysical Sciences, Princeton University, Princeton, NJ 08544}
\altaffiltext{2}{Physics Department, University of Illinois, Urbana, IL 61801}
\author{Scott C. Noble}
\affil{Center for Computational Relativity and Gravitation, School of Mathematical Sciences, Rochester Institute of Technology, 78 Lomb Memorial Dr, Rochester, NY 14623}

\email{jdolence@astro.princeton.edu}

\begin{abstract}

We report transient quasi-periodic oscillations (QPOs) on minute
timescales in relativistic, radiative models of the galactic center
source Sgr A*.  The QPOs result from nonaxisymmetric $m=1$ structure in
the accretion flow excited by MHD turbulence.  Near-infrared (NIR) and
X-ray power spectra show significant peaks at frequencies comparable to
the orbital frequency at the innermost stable circular orbit (ISCO)
$f_o$.  The excess power is associated with inward propagating magnetic
filaments inside the ISCO.  The amplitudes of the QPOs are sensitive to
the electron distribution function.  We argue that transient QPOs appear
at a range of frequencies in the neighborhood of $f_o$ and that the
power spectra, averaged over long times, likely show a broad bump near
$f_o$ rather than distinct, narrow QPO features.

\end{abstract}


\section{Introduction}

High frequency quasi-periodic oscillations (QPOs) in the light curves of
accreting black holes (e.g. \citet{Morg97,Stro01,Gier08}; see
\citet{Remi06} for a review) have attracted attention because they are a
potential probe of strong gravitational fields.  For example, QPO
frequencies may be related to $f_o$, the orbital frequency at the
innermost stable circular orbit (ISCO)---a feature unique to the strong
gravitational fields around black holes and possibly neutron stars.
Since in general relativity $f_o$ depends only on black hole mass $M$
and spin $a_*$, the QPO frequency, together with a mass estimate, could
be used to infer $a_*$.  But this cannot be done with confidence absent
a convincing QPO model.

It is difficult to discriminate between the many phenomenological and
physical models of QPOs \citep[e.g.][]{Remi06} using observations alone.
Numerical experiments offer a potential route to testing QPO models
\citep[e.g.][]{Kato04,Schn06,Chan09,Heni09}, but no numerical models
have yet produced robust high frequency QPOs.  \citet{Reyn09}, for
example, find no QPOs at all, while in other cases QPOs observed in
dynamical variables \citep{Heni09} are not observed in the emergent
radiation calculated from the same simulation \citep{Dext11}, and in
still other cases QPOs are observed only at certain times and viewing
angles \citep{Schn06}.

Still, accretion disks are complicated radiative and dynamical systems,
and faithfully modeling all the physics relevant to QPOs is a formidable
challenge.  In low luminosity systems such as Sgr A*, however, the
radiative and dynamical problems are decoupled and one might hope to
build more nearly {\em ab initio} models
\citep[e.g.][]{Mosc09,Dext09,Dext10,Mosc11,Dole11b}.  Searches for QPOs
in the near-infrared (NIR) and X-ray light curves of Sgr A* itself have
resulted in reported detections
\citep{Genz03,Asch04,Bela06,Ecka06,Meye06a,Meye06b,Trip07}, but these
may not be statistically significant \citep{Do09,Meye08}.  The
observational status of QPOs in Sgr A* is therefore unclear.

In this paper we present radiative models of Sgr A* based on three
dimensional GRMHD simulations.  We find clear evidence for QPOs in NIR
and X-ray light curves and power spectral densities (PSDs) on minute
timescales.  We tie the QPOs to $m=1$ structure excited by MHD
turbulence in the inner accretion flow.  We discuss the QPOs'
detectability and argue that their amplitudes are likely to be reduced
in Sgr A* relative to our simulations owing to the likely nonthermal
origin of the NIR emission.  We argue that the QPOs are transient and
appear over a range of frequencies near $f_o$ and therefore likely
result in a broad bump in PSDs averaged over sufficiently long
intervals.  Nonetheless, this is the first robust detection of a
radiative QPO in a self-consistent dynamical black hole accretion model.

\section{Model}

We model Sgr A* as a hot, optically thin, geometrically thick accretion
flow around a spinning black hole.  The disk orbital and black hole spin
angular momenta are aligned.   We set the dimensionless spin
$a_*=1-2^{-4}\simeq0.94$ based on axisymmetric models of the quiescent
spectrum \citep{Mosc09}.  Modeling proceeds in two stages: first we
evolve the disk using a GRMHD simulation, then we ``observe'' the disk
through relativistic radiative transfer calculations.

The fluid is evolved with the conservative three dimensional GRMHD code
{\tt harm3d} \citep{Nobl09,Gamm03}.  The computational volume extends
from within the horizon to $40\ml$ in radius $r$, [$0.02\pi$,$0.98\pi$]
in colatitude $\theta$, and [0,$2\pi$) in longitude $\phi$ with,
respectively, $192\times192\times128$ zones.  The zones are regularly
spaced in modified Kerr-Schild coordinates which are logarithmic in
radius and compressed near the equator to enhance resolution at small
radii and within the main body of the disk.  The initial conditions
consist of a quasi-equilibrium Fishbone-Moncrief torus with pressure
maximum near $12\ml$ and inner edge at $6\ml$, small perturbations to
the internal energy, and a weak purely poloidal magnetic field following
isodensity contours.  The disk is evolved for $\approx 11,500\mt$. After
an initial transient phase, at $t > 5000\mt$, the simulation data is
recorded every $0.5\mt$ for use in radiative transfer calculations.

The radiation field is evolved with the relativistic Monte Carlo code
\grm\ \citep{Dole09}.  \grm\ treats synchrotron emission and absorption
and Compton scattering.  It includes all relativistic effects including
finite light travel times through the time dependent GRMHD simulation
data.  Dimensional quantities are computed from the scale free GRMHD
simulation by specifying two numbers: $M = 4.5\times10^6\msun$
\citep{Ghez08,Gill09a,Gill09b}, and a disk mass unit $\mathcal{M}$
(effectively the accretion rate).  Following \citet{Dole11a}, we use a
smoothly varying time dependent $\mathcal{M}$ to remove long term trends
in the simulation data caused by draining of the initial disk onto the
hole.  

The radiation is recorded far from the hole and binned by photon
frequency $\nu$ in 38 ``cameras'' distributed quasi-uniformly over the
celestial sphere \citep[see][for a detailed discussion of how photons
are recorded]{Dole11c}.  The final data set includes broadband spectra
from radio to $\gamma$-rays in each camera with an integration time of
$\Delta=0.5\mt\approx11\,{\rm s}$ spanning
$\approx5100\mt\approx31.4\,{\rm hours}$ \citep[see][for
details]{Dole11b}.  The time-averaged spectra are in general agreement
with constraints from sub-mm VLBI, the sub-mm spectral slope, and limits
on the quiescent X-ray flux \citep{Mosc09} but underproduce NIR flux by
a factor of $\approx 10-30$ (see \S~\ref{sec:discussion})\footnote{See
\citet{Shio11} for a discussion of convergence of our GRMHD and
radiative simulations.}.

\section{Power Spectra}

The numerical data consists of light curves $L^\nu(t, \theta, \phi)$.
QPOs are strongest in the plane of the disk and absent when the disk is
face-on\footnote{QPOs are detectable simultaneously at all viewing
orientations except face-on.}.  For simplicity, then, we consider only
light curves at $\theta = \pi/2, \phi = k (2\pi/N_c)$, $0 \le k < N_c$,
recorded by the $N_c = 12$ equidistant cameras in the equatorial plane.
QPOs are detectable at all wavelengths $\lesssim100\mum$, but we
restrict attention to the NIR ($3.8\mum$ and $1.7\mum$) and X-ray
(integrated from $2-8$~keV) light curves since these are of greatest
interest for Sgr A*.

Following \citet{Pres92}, the light curves are divided into $N_{\rm
blocks}=19$ equal blocks.  We compute power spectral densities
(PSDs) for each of the $N_{\rm seg}=N_{\rm blocks}-1$ segments composed
of consecutive blocks and average these to obtain smoothed PSD
estimates.  We compute PSDs in both azimuthal wavenumber $m$ and in
temporal frequency $f_n = n /N_t\Delta$.  Before Fourier transforming,
the data is mean subtracted and Hamming windowed in time.  The discrete
Fourier transform is 
\begin{equation}\label{eq:transform} 
\tilde{L}^\nu_{mn} = \sum_{k=0}^{N_c-1} \sum_{l=0}^{N_t-1} 
	w_l (L^\nu_{kl}-\langle L^\nu\rangle) e^{-2\pi i(mk/N_c - nl/N_t)} \qquad 
	\left\{\begin{aligned} & -N_c/2 \le m \le N_c/2 \\ &
-N_t/2 \le n \le N_t/2 \end{aligned}\right.  
\end{equation} 
and the normalized PSD is 
\begin{equation}\label{eq:PSD} 
P(\nu;m,n) = \frac{1}{W_{ss}\langle L_\nu^2\rangle}\tilde{L}^\nu_{mn} \tilde{L}^{*\nu}_{mn}
\end{equation} 
where $W_{ss}=N_t\sum_{i=0}^{N_t-1} w_i^2$ is the window function
squared and summed as in \citet{Pres92} and $<L_\nu^2>$ is the mean
squared signal.  Since $L^\nu_{kl}$ is real we need only consider the
one-sided PSD defined at $f_n\ge0$.  With this choice $m>0$ ($m < 0$)
components circulate in the $+\phi$ ($-\phi$) direction.  Summing over
$m$ is equivalent to summing the PSDs from each camera, which is
justified by symmetry.  Averaging the PSD estimates over segments and
cameras improves the signal-to-noise-ratio.

\begin{figure}
\centering
\includegraphics[width=0.33\textwidth]{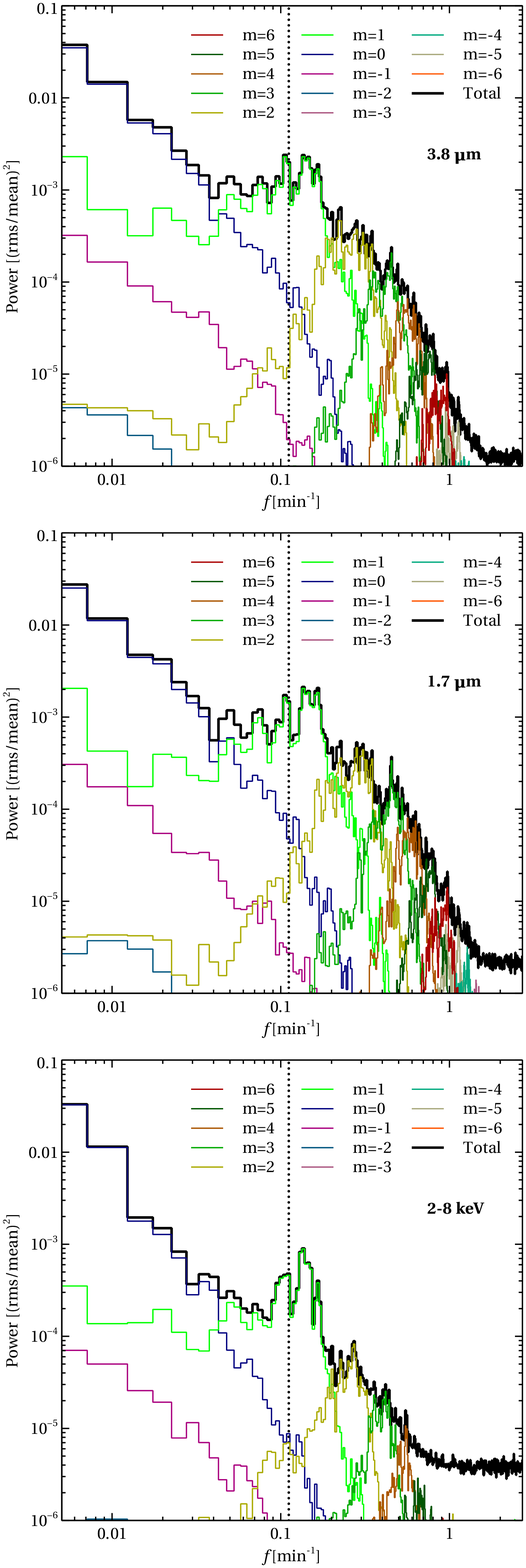}
\caption{Power spectra of light curves at $3.8\mum$ (top), $1.7\mum$ (middle), and integrated from 2--8$\kev$ (bottom).  QPOs are clearly seen near $f_o$ in all three power spectra, where the power is dominated by $m=1$ structure.  The dotted vertical line shows the ISCO frequency.}
\label{fig:spec}
\end{figure}

The normalized power spectra are shown in Fig.~\ref{fig:spec}.  The
total power is shown as a heavy solid line in the $3.8\mum$, $1.7\mum$,
and $2-8$keV bands, and the contributions from each $m$ are shown as
lighter colored lines.  The ISCO orbital frequency $f_o$ is shown as a
vertical dotted line.  The main result of this work is the existence of
several peaks superposed on a broad bump in the power spectrum near
$f_o$ in all three bands.  We identify these peaks as QPOs.

The PSDs show a power law dependence $P \sim f^{-2}$ at low frequency,
consistent with observations \citep[e.g.][]{Meye08,Do09}\footnote{Our
power spectrum does not sample frequencies low enough to see the PSD
break reported by \citet{Meye09}.}.  This low frequency component is
dominated by fluctuations in the $m = 0$ (axisymmetric) component of the
flow.  Near the ISCO frequency the $m = 1$ component dominates the
fluctuations and produces a broad, $Q\sim1$
($Q\equiv\nu/\Delta\nu_{FWHM}$), bump in the PSDs.  Superposed on this
broad feature are three peaks with centroids $f_1 = 0.106\imin$, $f_2 =
0.141\imin\,$, and $f_3 = 0.166\imin$ \footnote{$f_3 \simeq (3/2)f_1$,
but the significance of this is unclear}, which may be compared to $f_o
= 0.112\imin$.  The peaks have $Q \sim 10$.  At higher frequency $P
\sim f^{-3}$ and each $m>1$ contributes a broad peak near $f \sim m
f_o$.  At the highest frequencies the PSDs are dominated by white noise due to Monte Carlo shot noise.  The power in $m < 0$ components is negligible.

\section{Discussion}\label{sec:discussion}

Where do the QPOs originate?  The NIR and X-ray flux is dominated by
emission from $1.5\lesssim rc^2/GM \lesssim 2.5$ close to the equatorial
plane.  The QPO is therefore generated close to the ISCO, at $r_{ISCO}
\approx 2.044\ml$, and not far from the event horizon at $r_{hor}
\approx 1.348\ml$.  It clearly probes the strong gravitational field
regime.

The NIR flux is---in our model---thermal synchrotron emission from
electrons in the high energy tail of the distribution function.  The NIR
emissivity is sensitive to magnetic field strength and temperature,
which drop sharply with increasing radius.  The mean NIR emission is
therefore confined to a ring bounded at large radius by declining
emissivity and at small radius by gravitational redshift and photon
capture by the black hole\footnote{Optically thick disks, by contrast,
generate emission in a broad annulus at $\approx 2 r_{ISCO}$;  perhaps
this explains the lack of high frequency QPOs in the high, soft state in
black hole binaries.}.

The mean X-ray flux is dominated by synchrotron photons from the high
energy side of the synchrotron peak that are Compton upscattered once in
almost the same ring that generates the NIR emission.  This ring is
bounded at large radius by a decreasing probability of a large energy
amplification scattering due to decreasing temperature ($\Theta_e \equiv
k T_e/(m_e c^2) \sim r^{-3/2}$) and declining optical depth, and like
the NIR emission is bounded at small radius by gravitational redshift
and black hole photon capture.  Compton scattering occurs over a finite
radial range, however, and one would expect that finite light travel
time effects would tend to average away fluctuations on timescales less
than the light-crossing time.  This is consistent with
Fig.~\ref{fig:spec}, where the power at $f > f_o$ drops off more sharply
in the X-ray band than in the NIR: Compton scattering low-pass-filters
the NIR signal.

The fluctuating component of the NIR and X-ray flux need not arise in
the same place as the mean signal.  We investigated the origin of the
variable component by masking out emission from $r < r_{ISCO}$ and
recomputing segments of the light curve with a strong QPO signal.  In
these segments the QPO disappeared, confirming the importance of
sub-ISCO emission in generating the QPO.

\subsection{Camera footprint}

\begin{figure}
\centering
\includegraphics[width=0.8\textwidth]{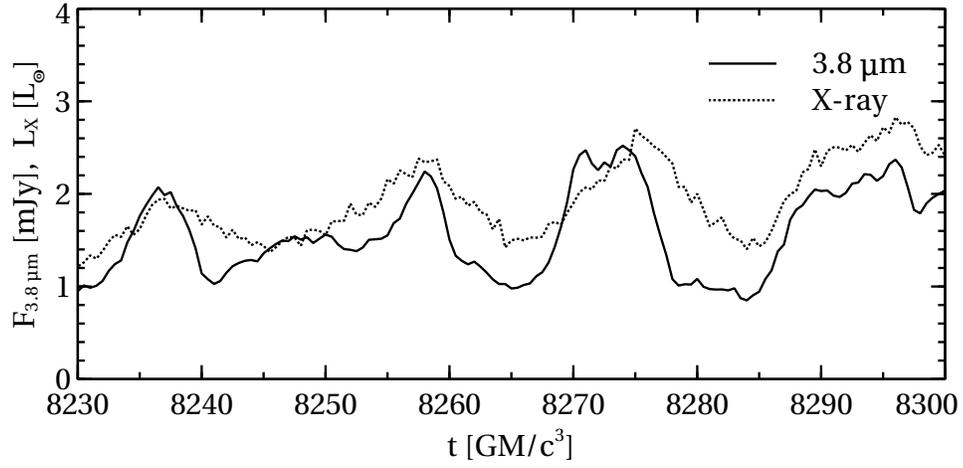}
\caption{NIR and X-ray light curves over a selected interval showing clear quasi-periodic structure.  Times shown correspond to the times of detection, $\sim 100\mt$ after emission.}
\label{fig:lc}
\end{figure}

\begin{figure}
\centering
\includegraphics[width=0.8\textwidth]{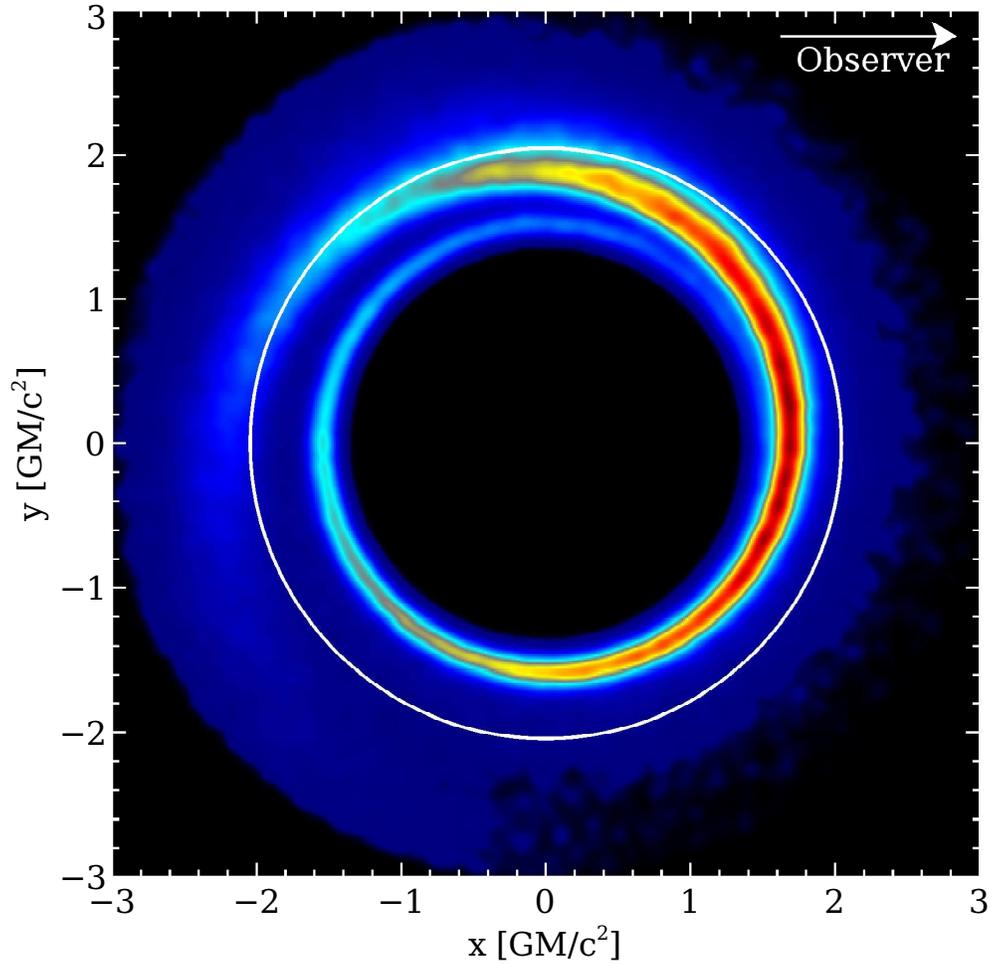}
\caption{Time-averaged distribution function $dN/dxdy$ for the origin of NIR photons detected by an observer far away along the $+x$-axis.  The black hole spin and disk orbital motion are counterclockwise and the ISCO is indicated by the solid white line.}
\label{fig:map}
\end{figure}

Because NIR and X-ray photons are generated so close to the event
horizon, gravitational lensing, Doppler beaming, and time delays play a
role in generating the observed signal.  To investigate the geometry of
the emission region we have focused on a small segment of the light
curve where a QPO is particularly strong.  Part of this curve is shown
in Fig.~\ref{fig:lc}.  Within the segment we generated a vertically
integrated map of the radius and azimuth (relative to the observing
camera) of the origin of photons that are detected in the NIR.  The
resulting map of $dN/dxdy$ is shown in Fig.~\ref{fig:map}.  The disk
orbits counterclockwise around the hole and the camera is at the far
right.  Evidently each camera detects emission from a region that forms
a slender, {\em leading} spiral around the hole.  The spiral footprint
extends for almost $3\pi$ radians.

\begin{figure}
\centering
\includegraphics[width=0.8\textwidth]{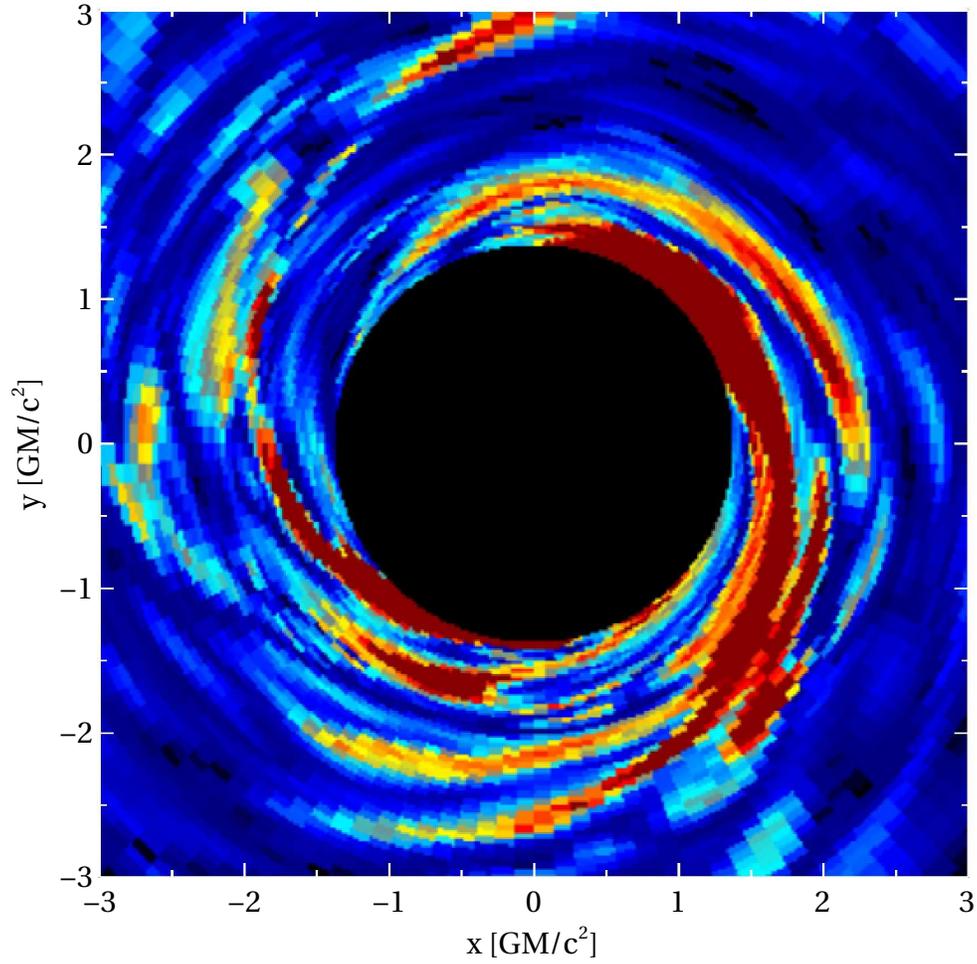}
\caption{A snapshot at $t=8120\mt$ showing $\rho B^2$ in the equatorial plane clearly indicating the presence of $m=1$ structure.}
\label{fig:rhob2}
\end{figure}

The camera footprint is radially narrow but azimuthally extended.  Since
the observation process convolves disk structure with the camera
footprint, the footprint filters out fluctuations from azimuthally
narrow structures (evident in Fig.~\ref{fig:rhob2}; indeed most studies
show flat spatial power spectra for disk turbulence from $m = 1$ up to
$m \sim R/H \times$ a few).  The decline in power above $\sim 2 f_o$
(and thus, according to Fig.~\ref{fig:spec}, with increasing $m$) is
partially due to this smoothing effect of the camera footprint.  The
radial narrowness of the footprint also implies sensitivity to radially
narrow structures.  Radial infall of an axisymmetric emitting region
through the footprint would yield variability.  No nonaxisymmetric
structure is required.  Nevertheless, our analysis shows that the QPOs
stem from $m = 1$ structure in the emitting plasma.

\subsection{Underlying flow structure}

What is the $m = 1$ structure that generates the QPOs?  We investigate
this by suppressing nonaxisymmetric structure in each fluid variable in
turn and recalculating a segment of the light curve with a strong QPO
signal.  This procedure reveals that the QPOs are mainly generated by
variations in magnetic field {\em strength}\footnote{Using an
angle-averaged emissivity barely alters the QPO amplitude, so variations
in field {\em direction} are not essential.}; Fluctuations in the
temperature contribute, but at a lower level.  Lightcurves in which
nonaxisymmetric structure in the magnetic field and temperature are
removed show little variability.  

If we ignore the dips in the power spectra between the peaks at $f_1$
and $f_3$, the QPO is a broad feature with $Q \sim 1$.  The individual
peaks at $f_1$, $f_2$, and $f_3$ have $Q \sim 10$.  On average they are,
therefore, not very long-lived or coherent structures, as one might
expect if they are due to turbulence in the disk.  They are {\em not}
orbiting blobs; we see no evidence for coherent, orbiting ``hot spots''
in the disk.  

Figure~\ref{fig:rhob2} shows a single snapshot of $\rho b^2$ in the
midplane close to the black hole, which shows that there is large-scale
structure in the magnetic field.  Visual inspection of animations of
similar images shows that trailing spiral magnetic filaments propagate
inward and move approximately with the fluid velocity.  That the
QPO-generating features track the fluid velocity strongly suggests that
the QPO frequency should scale with the orbital frequency at the ISCO,
and therefore vary with $a_*$, but we have not yet analyzed models with
different spin.

\subsection{Significance and persistence of individual peaks}

\begin{figure}
\centering
\includegraphics[width=0.33\textwidth]{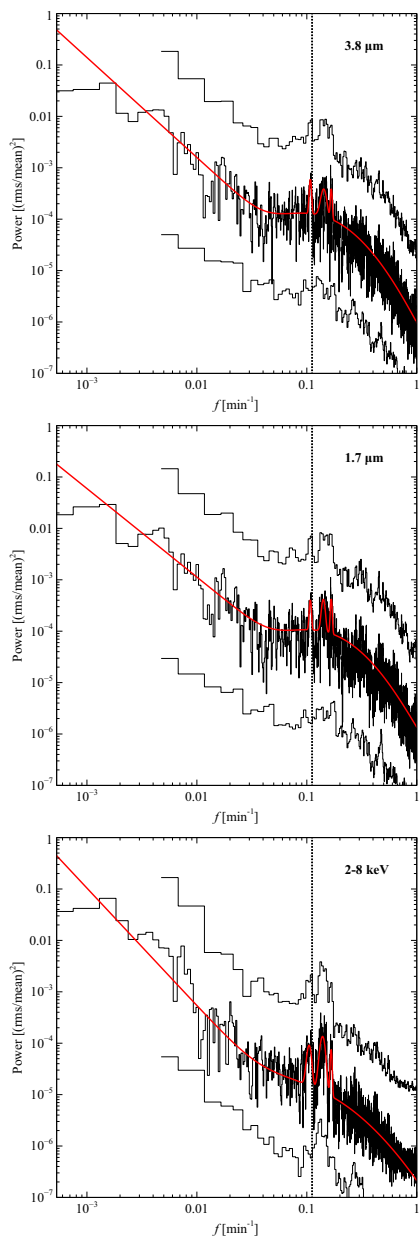}
\caption{The noisy central line in each panel shows the PSDs from the full light curves without averaging over segments.  The smooth lines show the best fit models.  The bottom and top lines in each panel show the PSDs from the first and last half of the simulation, respectively, averaged over segments as in Fig.~\ref{fig:spec}, renormalized to account for the lower frequency resolution, and then offset by factors of 30.}
\label{fig:raw}
\end{figure}

Are our QPOs statistically significant?  To check, we compute PSDs of
the full unsegmented light curves and fit each with a sequence of nested
models including from zero to three QPOs.  Likelihood-ratio tests then
indicate that all three QPOs are highly significant in all three power
spectra, with p-values $\lesssim5\times10^{-8}$.  The fitted PSDs and
the best-fit models are shown in Fig.~\ref{fig:raw}.

Though significant, the QPOs are not persistent.  Figure~\ref{fig:raw}
shows the PSDs from the first and last half of the simulation, averaged
over segments as in Fig.~\ref{fig:spec}.  We see that the PSDs from the
full simulation are very similar to those from the last half alone.
This is because the last half of the simulation includes weak flares
that dominate the total power.  We also see that the PSDs from the first
half show QPO-like features but with slightly different frequencies,
though close to $f_o$.  This suggests that PSDs computed from longer
time series would have broad $Q\sim1$ bumps near $f_o$, but finite
length realizations of the light curves are likely to show narrower QPO
signals at one or more frequencies that may reflect real quasi-periodic
behavior.  Our model shows transient QPOs with $Q\sim10$ that appear
stochastically at frequencies near $f_o$.

\subsection{Detectability}

Are the QPOs detectable?  Recall that our model underproduces the NIR
flux of Sgr A* by a factor of 10--30.  Also, the observed $\nu
F_\nu\propto\nu^{0.5}$ \citep{Horn07,Trap11}, whereas our model predicts
a red spectral slope.  This can be understood if the electron
distribution function has a quasi-thermal core and a nonthermal,
high-energy tail.  For a power-law distribution of electrons with number
density $n_e$, $dn_e/d\gamma \sim \gamma^{-p}$, the emissivity is
$\propto \nu^{-(p - 1)/2}$, so the observed NIR slope implies $p \simeq
2$, a common slope for synchrotron-emitting sources.  The change in NIR
emissivity may affect the detectability of the QPO.  

We have not yet incorporated a nonthermal tail of electrons into our
time-dependent model \citep[see][where time-averaged models with a
nonthermal tail can produce Sgr A*'s NIR flux]{Leun10}.  Naively, one
would expect a nonthermal model to be less sensitive to $B$ than a
thermal model: a nonthermal component with fixed $n_e$ has $j_\nu
\propto B^{(p + 1)/2}$, i.e. $d\ln j_\nu/d\ln B \simeq 1.5$, whereas in
the NIR our thermal model has $j_\nu \propto \exp(-(\nu/\nu_s)^{1/3})$,
where $\nu_s(\Theta_e,B)$ is a characteristic frequency for synchrotron
emission.  For parameters appropriate to the NIR in Sgr A*, $d\ln
j_\nu/d\ln B \simeq 3$, suggesting a reduced amplitude for a
nonthermally generated QPO.  On the other hand, the density of
nonthermal electrons may be sensitive to $B$ and $\Theta_e$.  That some
nonlinear sensitivity of the emissivity to $B$, $\Theta_e$, or $n_e$ is
required is consistent with the absence of a QPO near the sub-mm peak,
where the emissivity is only weakly sensitive to all three.  An accurate
assessment of QPO strength for any particular nonthermal component will
require a full, time-dependent model.  

Nonetheless, a prominent feature in the power spectrum close to $f_o$ in
geometrically thick, optically thin accretion flows seems robust.  Hot
flows concentrate their emission in a narrow ring near the ISCO whereas
optically thick disks do not.  The emission is (strongly or weakly)
variable due to the inevitable presence of turbulence in the accretion
flow, and variation at frequencies $f \gg f_o$ due to small scale
structure will inevitably be averaged away by the observing process.

\section{Summary}

We have performed GRMHD and radiative simulations of the accretion flow
in Sgr A* and found prominent features near the ISCO orbital frequency
in both NIR and X-ray light curves.  These features are not present for
face-on observers or in the millimeter or submillimeter.  We have shown
that: (1) the features have an $m = 1$ structure on the celestial sphere
as seen from the source, and are therefore due to $m = 1$ structure in
the source; therefore (2) full $2\pi$ azimuthal domain models are
required to accurately model the light curves of similar sources; (3)
the variable emission arises near and inside the ISCO and therefore
probes a strongly relativistic regime close to the event horizon; (4)
observations in the NIR and X-ray bands are sensitive to a narrow spiral
footprint on the disk midplane; (5) the variability, for our emission
model, is dominated by variations in the magnetic field strength, with a
lesser contribution from variations in the disk temperature; (6) the
varying features move approximately with the fluid velocity, therefore
(7) the centroid frequencies should be sensitive to the ISCO orbital
frequency and therefore black hole spin.  Discovery of similar features
in the variability spectrum of Sgr A* would be an exciting opportunity
to probe the spin of the galaxy's central black hole.

\acknowledgements

This work was supported by NASA under NASA Earth and Space Science
Fellowship NNX10AL24H for JCD, the National Science Foundation under
grant AST 07-09246, NASA under grant NNX10AD03G, the NSF through
TeraGrid resources provided by NCSA and TACC, and by a Richard and
Margaret Romano Professorial scholarship, and a University Scholar
appointment to CFG.  Part of this work was completed while CFG was a
visitor at Max-Planck-Institut f\"ur Astrophysik. CFG thanks Henk Spruit
and Rashid Sunyaev for their hospitality.

\newpage

\newpage

\newpage

\newpage

\newpage

\end{document}